\documentclass[10pt,twoside]{hsqcd}
 \usepackage{epsf,amsmath}
  \usepackage[dvips,final]{graphicx}
   \usepackage{amssymb}
    \usepackage{amsmath}
     \usepackage{pifont}
\begin{document}
\title{{\hfill RUB-TPII-04/04}\\ [1cm] Pion structure: from nonlocal
        condensates to NLO analytic perturbation theory\thanks{Invited
        plenary talk presented by the first author at \textit{Hadron
        Structure and QCD: from Low to High Energies}, St. Petersburg,
        Repino, Russia, 18-22 May 2004.}}
\author{N.~G.~Stefanis$^{1,a}$, A.~P.~Bakulev$^{2,b}$, S.~V.~Mikhailov$^{2,c}$,\\
        K.~Passek-Kumeri\v{c}ki$^{3,d}$, W.~Schroers$^{4,e}$ \\
$^1$Institut f\"ur Theoretische Physik II,
Ruhr-Universit\"at Bochum, \\
D-44780 Bochum, Germany \\
$^a$Email: stefanis@tp2.ruhr-uni-bochum.de\\
$^2$Bogoliubov Laboratory of Theoretical Physics, JINR,
141980 Dubna, Russia\\
$^b$Email: bakulev@thsun1.jinr.ru \\
$^c$Email:mikhs@thsun1.jinr.ru \\
$^3$Theoretical Physics Division, Rudjer-Bo\v{s}kovi\'{c}
Institute P.O.\ Box 180, \\ HR-10002 Zagreb, Croatia \\
$^d$Email: passek@thphys.irb.hr \\
$^4$Center for Theoretical Physics, Laboratory for Nuclear Science \\
and Department of Physics, MIT, Cambridge Massachusetts 02139,
USA\\
$^e$Email: wolfram.schroers@feldtheorie.de}

\maketitle
\bigskip
\begin{abstract}
\noindent A pion distribution amplitude, derived from nonlocal QCD
sum rules, has been employed to calculate
$F_{\gamma^*\gamma\to\pi}(Q^2)$ using light-cone sum rules, and
$F_{\pi}(Q^2)$ in NLO QCD perturbation theory. Predictions are
presented for both observables and found to be in good agreement
with the corresponding data. Calculating the hard pion form factor
by Analytic Perturbation Theory to two-loop order, it is shown
that the renormalization-scheme and scale-setting dependencies are
diminished.
\end{abstract}

\markboth{\large \sl N.~G.~Stefanis et al.  \hspace*{2cm} HSQCD
2004} {\large \sl \hspace*{1cm} Pion structure: from nonlocal
condensates to NLO APT}

\section{Introduction}
Large-distance QCD remains an area, where the concepts of
perturbation theory cannot be directly applied. To assess this
region and make reliable predictions for hadronic processes, the
pure perturbative treatment has to be amended by nonperturbative
input.

In a series of recent papers \cite{BMS01,BMS02}, three of us have
outlined an approach, based on QCD sum rules with nonlocal
condensates \cite{MR86}, capable of providing a pion distribution
amplitude (DA) compatible at $1\sigma$ with the CLEO data
\cite{CLEO98} on the pion-photon transition. The key feature of
this pion DA is that its endpoint regions $x=0,1$ ($x$ being the
parton's longitudinal momentum fraction) are strongly suppressed.
This suppression is controlled by the nonlocality of the scalar
quark condensate, parameterized by the average quark virtuality
$\lambda_q^2$ in the vacuum, with theoretical estimates in the
range $(0.4 - 0.5)~\mbox{GeV}^2$ \cite{BM02} and a preferable
value of $0.4$~GeV$^2$ extracted in \cite{BMS02} from the CLEO
data.

In addition, one can improve the quality of perturbatively
calculable observables, notably the factorized hard contribution
of the pion's electromagnetic form factor, by trading the
traditional power-series perturbative expansion for a
non-power-series (in an analytic) coupling expansion that avoids
{\it eo ipso} the Landau singularity rendering all expressions
infrared (IR) finite \cite{SS97,Shi-an}. Suffice it here to say
that this is achieved through the inclusion into the running
coupling of a power-behaved term of nonperturbative origin that
removes the Landau ghost leaving the ultraviolet behavior of the
effective coupling unchanged. Crucial for making this analytic
approach possible, is the generalization of the analytic
running-coupling concept, proposed by Shirkov and Solovtsov
\cite{SS97}, to the level of observables depending on more than
one scheme scales \cite{KS01}, as is, for example, the case for
the pion form factor in fixed-order perturbation theory beyond LO
\cite{BPSS04}, or performing a Sudakov resummation \cite{SSK99}.

Along these lines of thoughts, we describe in this contribution
our recent works on the pion DA, summarizing the main results, and
present predictions for the pion's electromagnetic form factor
carried out under the imposition of analyticity of the running
coupling and its powers using two different procedures. It turns
out that if the powers of the coupling have their own analytic
(dispersive) images, the factorizable hard part of the form factor
so calculated bears a minimal dependence on the scheme and
scale-setting choice. Including also the soft contribution via
local duality, this helps improving the quality of the prediction
beyond the level of the current experimental-data accuracy.

\section{Endpoint-suppressed pion distribution amplitude}
In the context of factorization of hard exclusive processes
\cite{ERBL79}, the pion DA is a universal, gauge-invariant
quantity defined at the twist-2 level by
\begin{equation}
  {\langle 0\mid \bar{d}(z) \gamma^{\mu}\gamma_{5} {\cal C}(z,0) u(0)
            \mid \pi(P)
  \rangle}\Big|_{z^2=0}
=
  i f_{\pi} P^{\mu} \int^{1}_{0} dx {\rm e}^{ix(zP)}
  \varphi_{\pi}\left(x,\mu_{0}^{2} \sim z^{-2}\right)
\label{eq:def-phi-pi}
\end{equation}
where $\int^{1}_{0} dx \varphi_{\pi}\left(x,\mu^{2}\right)=1$,
$f_{\pi} = 131$~MeV is the pion decay constant, and
$
  {\cal C}(0,z)
  = {\cal P}
  \exp\!\left[-ig_s\!\!\int_0^z t^{a} A_\mu^{a}(y)\, dy^\mu\right]
$
preserves gauge invariance. $\varphi_\pi(x,\mu^2)$ encapsulates
the nonperturbative QCD pion structure in terms of the
distribution of the longitudinal momentum fractions between its
two valence partons: quark ($x$) and antiquark ($\bar{x}\equiv
1-x$). Together with the DA of its first resonance, $A_1$, it can
be related to the nonlocal condensates by means of a sum rule,
based on the correlator of two axial currents (see \cite{BMS01}).
Due to the finiteness of the vacuum correlation length
$\lambda_q^{-1}$, the end-point regions $x\to 0, 1$ are strongly
suppressed and by virtue of this fact we can \cite{BMS01}
determine quite accurately the first ten moments $
 \langle\xi^N\rangle_\pi
\equiv \int_{0}^{1}
 \varphi_\pi(x)(2x-1)^N dx
$
of the pion DA and {\it independently} also the inverse moment
$
 \langle x^{-1}\rangle_{\pi}
\equiv
 \int_{0}^{1}\varphi_\pi(x)\ x^{-1} dx
$. Given that $\langle \xi^N \rangle_\pi\to \langle \xi^N
\rangle^{as}_\pi$ rapidly with increasing $N$, the eigenfunctions
decomposition
\begin{eqnarray}
 \varphi_\pi(x,\mu_{0}^{2})
  = 6 x (1-x)
     \left[ 1
          + a_2(\mu_{0}^{2}) \, C_2^{3/2}(2 x -1)
          + a_4(\mu_{0}^{2}) \, C_4^{3/2}(2 x -1)
          + \ldots
     \right]
\label{eq:phi024mu0}
\end{eqnarray}
can be practically truncated at $a_4$ because all higher
coefficients are negligible \cite{BMS01}. The ``bunch'' of the
pion DAs shown in Fig. 1(a), parameterized by $a_2$ and $a_4$,
turns out to match all moment constraints for $\langle
\xi^{N}\rangle_{\pi}$ and $\langle x^{-1}\rangle_{\pi}$ extracted
from the CLEO data. The optimum sample out of this ``bunch''---BMS
model---\cite{BMS01}, has at $\mu_0=1~{\rm GeV}$ $a_2=0.20$ and
$a_4=-0.14$ and is shown in Fig.\ 1(a). Let us close this section
with a forward-looking statement: the BMS ``bunch'' pion DAs,
though doubly peaked, have their endpoints ($x\to 0,1$) strongly
suppressed---not only relative to $\varphi_{\pi}^{\rm CZ}$ but
even compared to $\varphi_{\pi}^{\rm as}$, substantially reducing
the importance of Sudakov effects.
\begin{figure}[t]
 \centerline{\includegraphics[width=0.47\textwidth]{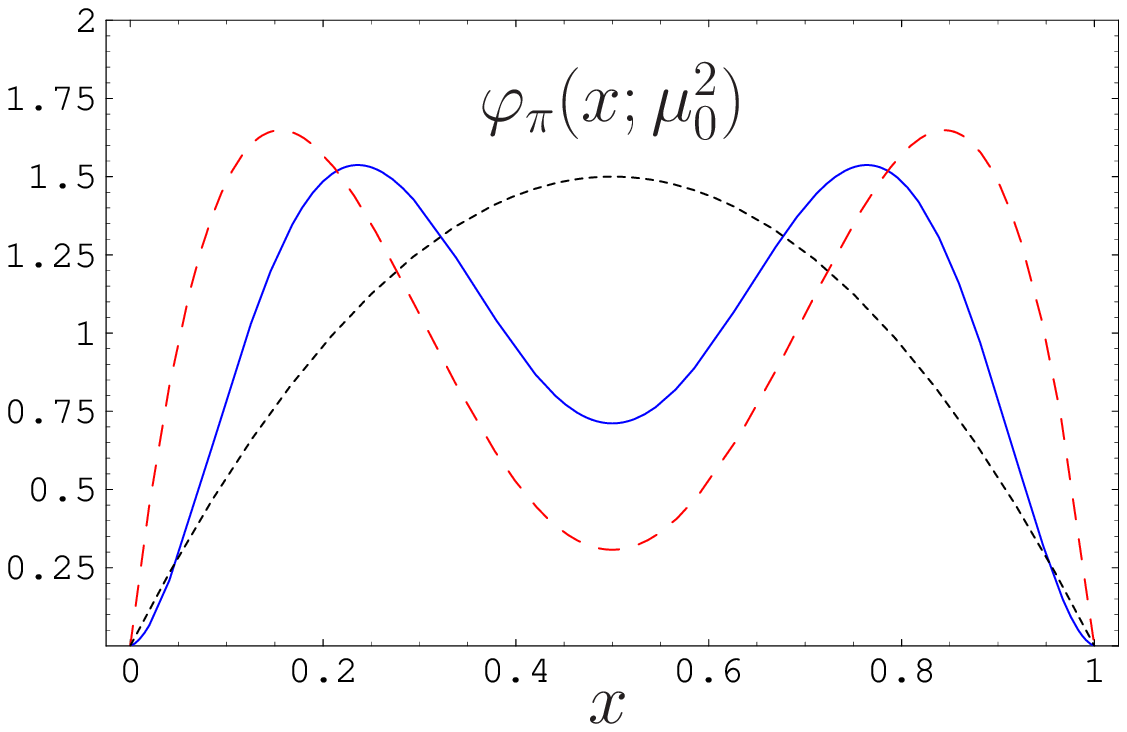}~~~%
             \includegraphics[width=0.47\textwidth]{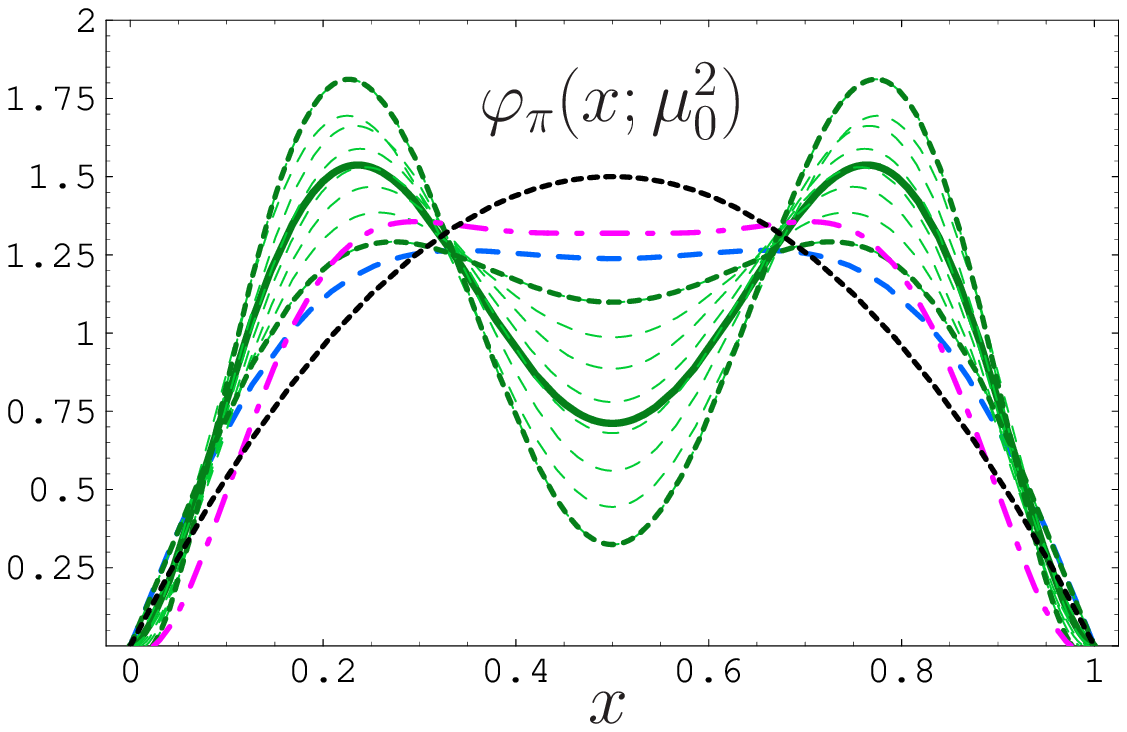}}
   \caption[*]{(a) Profiles of pion DAs normalized at
   $\mu_0^2 = 1~{\rm GeV}^2$:
   BMS model~\protect\cite{BMS01}---solid line;
   CZ model~\protect\cite{CZ84}---dashed line;
   asymptotic DA---dotted line.
   (b) BMS ``bunch'' \protect\cite{BMS01} in comparison with
   $\varphi_\text{as}$,
   $\varphi_\text{PR}$ (dashed line) \protect\cite{PR01},
   $\varphi_\text{Dor}$ (dash-dotted line) \protect\cite{Dor02}.
\label{fig:pi-DAs}}
\end{figure}

\section{Comparison with the CLEO data on the pion-photon
transition} It was shown in \cite{Kho99} at LO and later extended
\cite{SY99} to NLO of QCD perturbation theory \cite{DaCh81} that
the light-cone QCD sum-rule method allows to perform all
calculations in the $\gamma^{*}(Q^{2})\gamma(q^{2})\to\pi^{0}$
form factor for sufficiently large $q^{2}$ and analytically
continue the results to the limit $q^{2}=0$, hence avoiding
problems arising when a photon becomes real. Recently
\cite{BMS02}, we have revised and refined this sort of data
processing accounting for a correct ERBL \cite{ERBL79} evolution
of the pion DA, 
including thresholds effects in the running coupling, estimating
more accurately the contribution of the twist-4 contribution, and
improving the error estimates in determining the $1\sigma$- and
$2\sigma$-error contours. Avoiding here technical details, we
gather the results of our analysis in Fig. 2. The predictions
shown correspond to the following pion DA models with associated
$\sigma$ deviations and designations for the form-factor
predictions displayed in the right panel: $\varphi_{\rm CZ}$
({\footnotesize\ding{110}}, $4\sigma$, upper dashed line)
\cite{CZ84}; BMS-``nonlocal QCD SRs bunch'' (shaded rectangle),
$\varphi_{\rm BMS}$ (\ding{54}, $1\sigma$---left panel, shaded
strip---right panel) \cite{BMS01}; three instanton-based models,
viz., \cite{PPRWG99} (\ding{72}, $3\sigma$, dotted line),
\cite{PR01} (\ding{70}, $2\sigma$, dash-dotted line), and
\cite{ADT00} ($\blacktriangle$, $3\sigma$---only left panel); and
the asymptotic pion DA $\varphi_{\rm as}$ (\ding{117}, $3\sigma$,
lower dashed line). A recent transverse lattice result
\cite{Dal02} ({\footnotesize\ding{116}}, $2\sigma$) is also
shown---left panel only.

\begin{figure}[thb]
 \centerline{\includegraphics[width=0.47\textwidth]{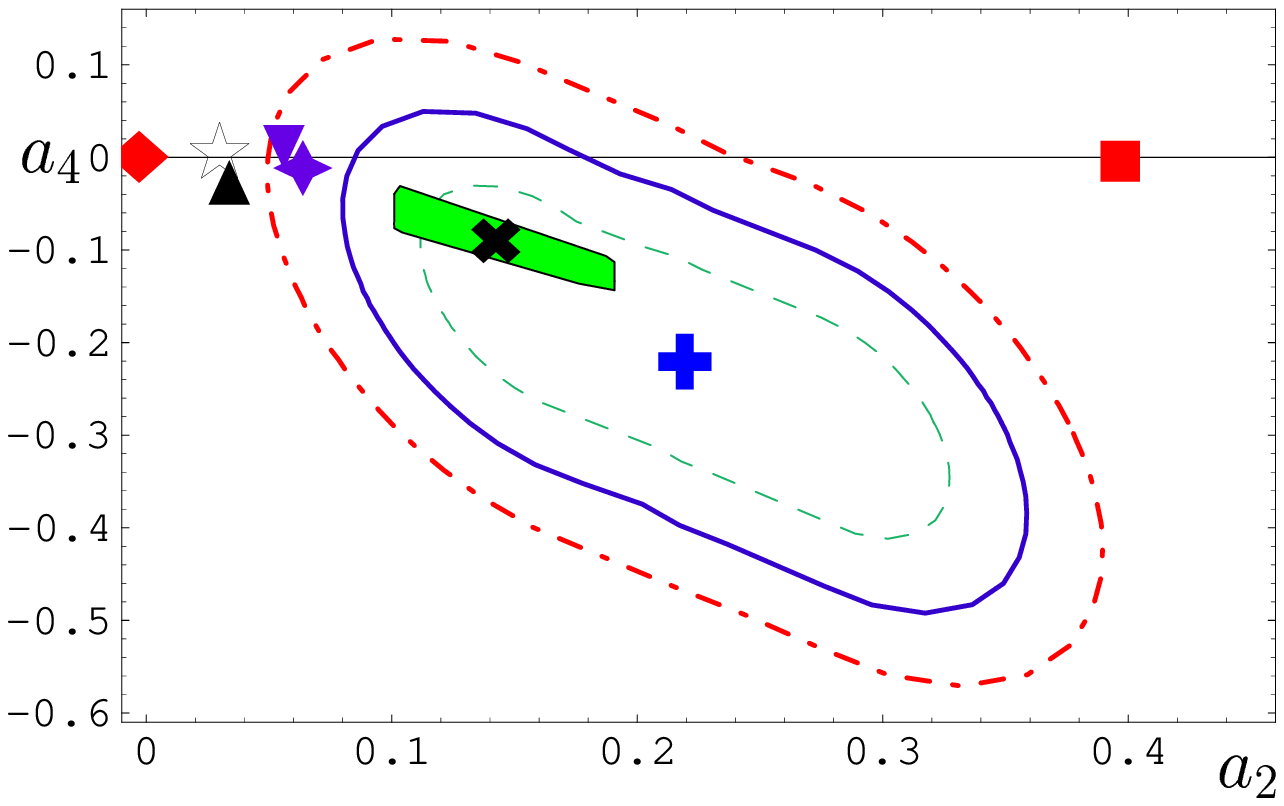}~~~%
             \includegraphics[width=0.47\textwidth]{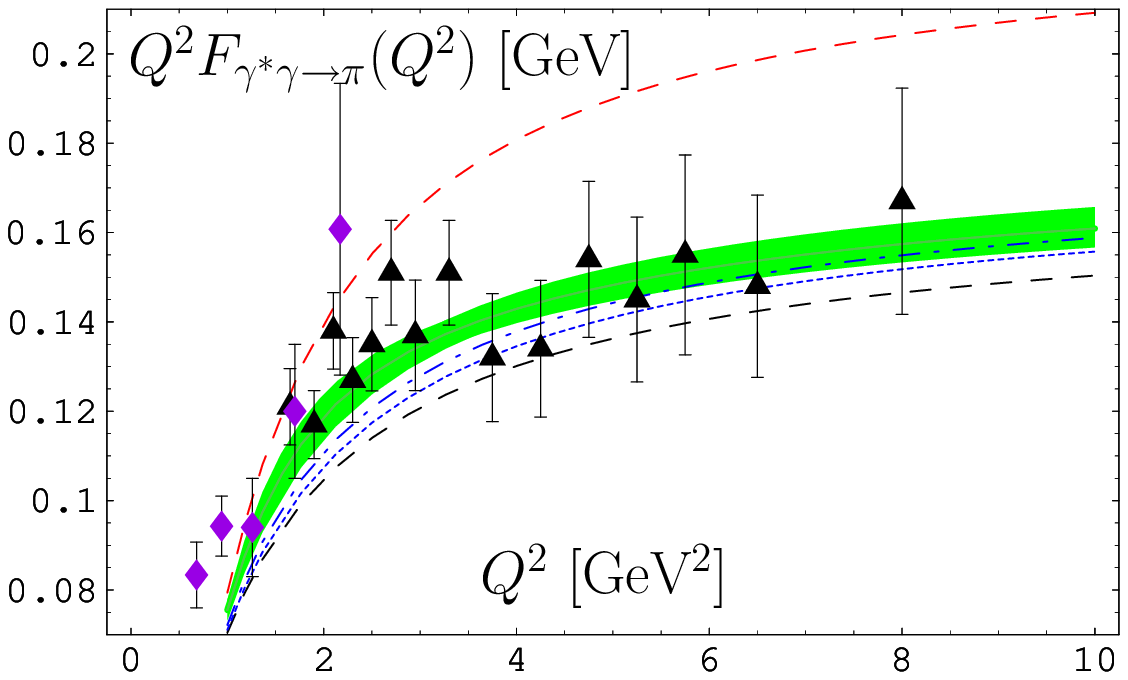}}
  \caption[*]{(a) $Q^2F_{\gamma^*\gamma\to\pi}(Q^2)$ CLEO-data analysis
   in terms of error contours in the ($a_2$,$a_4$) plane.
   The line assignments are:
   broken line---$1\sigma$; solid line---$2\sigma$; dash-dotted
   line---$3\sigma$.
   Various pion DAs are shown, evaluated at
   $\mu^{2}_{\rm SY}=5.76$~GeV$^2$ after NLO ERBL evolution..
   The slanted shaded rectangle represents the nonlocal QCD sum-rule
   constraints on ($a_2,~a_4$) \protect\cite{BMS01} for
   $\lambda^2_{\rm q}=0.4$~GeV$^{2}$.
   (b) Light-cone sum-rule predictions for
   $Q^2F_{\gamma^*\gamma\to\pi}(Q^2)$
   in comparison with the CELLO (diamonds, \protect\cite{CELLO91})
   and the CLEO (triangles, \protect\cite{CLEO98}) data
   evaluated with
   $\delta_{\rm Tw-4}^2=0.19$~GeV$^2$ \protect\cite{BMS02}.
\label{fig:pi-photonFF}}
\end{figure}
To summarize, the main results obtained in \cite{BMS02} are: (i)
Both DAs, $\varphi_{\pi}^{\rm as}$ \cite{ERBL79} and
$\varphi_{\pi}^{\rm CZ}$ \cite{CZ84} are disfavored by the CLEO
data at $3\sigma$ and $4\sigma$, respectively. In contrast,
$\varphi_{\pi}^{\rm BMS}$ lies within the $1\sigma$-error ellipse.
Model DAs from instanton-based approaches
\cite{PR01,Dor02,PPRWG99,ADT00} are close to but still outside the
$2\sigma$ region. (ii) The extracted  coefficients $a_{2}$ and
$a_{4}$ are rather sensitive to the strong radiative corrections
and the size of the twist-4 contribution. (iii) The value of the
vacuum nonlocality extracted from the CLEO data is $\lambda_{\rm
q}^{2}\lesssim 0.4$~GeV$^{2}$. Turning to the form-factor
predictions, one observes from Fig.\ 2 (Right) that the BMS
``bunch'' of pion DAs is in good agreement with the CLEO data
\cite{CLEO98} but also with the CELLO data \cite{CELLO91}, while
the behavior of rival DAs reflects the situation shown in the left
panel: the prediction from the CZ model overshoots the data
considerably, while that from $\varphi_{\pi}^{\rm as}$---and DAs
close to it---are underestimating both sets of experimental data.

\section{Electromagnetic pion form factor. Theory and phenomenology}
The crucial new elements of the calculation below are: (i) use of
the BMS pion DA, (ii) application of two-loop Analytic
Perturbation Theory (APT), and (iii) a more accurate way, based on
local duality, to join the soft part with the hard form-factor
contribution.

The pion's electromagnetic form factor can be generically written
as \cite{ERBL79}
$
    F_{\pi}(Q^{2})
=   F_{\pi}^{\rm Fact}(Q^{2})
  + F_{\pi}^{\rm non-Fact}(Q^{2}),
$ where $F_{\pi}^{\rm Fact}(Q^{2})$ is the factorized part within
pQCD and $F_{\pi}^{\rm non-Fact}(Q^{2})$ is the ``soft'' part
containing subleading power-behaved (e.g., twist-4) contributions
originating from nonperturbative effects. The leading-twist
factorizable contribution can be expressed as a convolution in the
form
$
F_{\pi}^{\rm Fact}(Q^{2}; \mu_{\rm R}^{2}) =
  f_{\pi}^{2}\varphi_{\pi}^{*}(x,\mu_{\rm F}^{2})
  \otimes
  T_{\rm H}(x,y,Q^{2};\mu_{\rm F}^{2},\mu_{\rm R}^{2})
  \otimes
  \varphi_{\pi}(y,\mu_{\rm F}^{2})
$, where $\mu_{\rm F}$ is the factorization scale between the
long- and short-distance dynamics, $\mu_{\rm R}$ stands for the
renormalization scale. The hard-scattering amplitude, $T_{\rm
H}(x,y,Q^{2};\mu_{\rm F}^{2},\mu_{\rm R}^{2})$, describing
short-distance interactions at the parton level, has been
evaluated to NLO accuracy (\cite{MNP99a} and references cited
therein) using the terminology introduced in \cite{BPSS04} to
which we refer for details. Then, one obtains $ F_{\pi}^{\rm
Fact}(Q^2; \mu_{\rm R}^{2})
  =  F_{\pi}^{\rm LO}(Q^2;   \mu_{\rm R}^{2})
   + F_{\pi}^{\rm NLO}(Q^2;  \mu_{\rm R}^{2})
$,
where the LO and NLO terms read, respectively,
\begin{eqnarray}
  F_{\pi}^{\rm LO}(Q^2;\mu_{\rm R}^{2})
&=&
  \alpha_{\rm s}(\mu_{\rm R}^{2})\, {\cal F}_{\pi}^{\rm LO}(Q^2)\\
  Q^2 {\cal F}_{\pi}^{\rm LO}(Q^2)
&\equiv&
  8\,\pi\,f_{\pi}^2\,
    \left[1 + a_2^{\rm D,NLO}(Q^2) + a_4^{\rm D,NLO}(Q^2)\right]^2 \, ,
\label{eq:Q2pffLO}
\end{eqnarray}
\begin{equation}
  F_{\pi}^{\rm NLO}(Q^2;\mu_{\rm R}^{2})
=
    \frac{\alpha_{\rm s}^2(\mu_{\rm R}^{2})}{\pi}\,
    \left[{\cal F}_\pi^{\rm D,NLO}(Q^2;\mu_{\rm R}^{2})
  + {\cal F}_\pi^{\rm ND,NLO}(Q^2;N_{\rm Max}=\infty) \,
    \right].
\label{eq:Q2pffNLO}
\end{equation}
Here $N_{\rm Max}$ marks the maximal number of Gegenbauer
harmonics taken into account and the calligraphic designation
denotes quantities with their $\alpha_{\rm s}$-dependence pulled
out. Note that because we take into account the NLO evolution of
the pion DA, the displayed terms contain diagonal (D) as well as
(the NLO term) non-diagonal (ND) components. The effects of the LO
DA evolution are crucial \cite{MNP99a}, while the NLO ones are
relatively of less importance. Hence, we set here:  $
  a_n^{\rm D,NLO} \to a_n^{\rm D,LO}
 \;\, \mbox{and} \;\,
  a_n^{\rm ND,NLO} \to  0.\,
$
Studying $F_{\pi}$ beyond the LO requires an optimal
renormalization scheme and scale setting in order to minimize the
influence of higher-order loop corrections (see \cite{BPSS04} for
a fully fledged discussion). To join the hard with the soft
contribution (the latter being calculated with the aid of local
duality (LD), we have to correct the low-$Q^2$ behavior of the
factorizable part to fulfill the Ward identity at $Q^2=0$, i.e.,
$
F_{\pi}(Q^{2};\mu_{\rm R}^{2}) =  F_{\pi}^{\rm LD}(Q^{2})
  +  F_{\pi}^{\rm Fact-WI}(Q^2;\mu_{\rm R}^{2})\,
$ with
$
 F_{\pi}^{\rm Fact-WI}(Q^2;\mu_{\rm R}^{2})
= Q^4/\left(2s_0^{\rm 2-loop}+Q^2\right)^2
       F_{\pi}^{\rm Fact}(Q^2;\mu_{\rm R}^{2}).
$
\begin{figure}[thb]
 \centerline{\includegraphics[width=\textwidth]{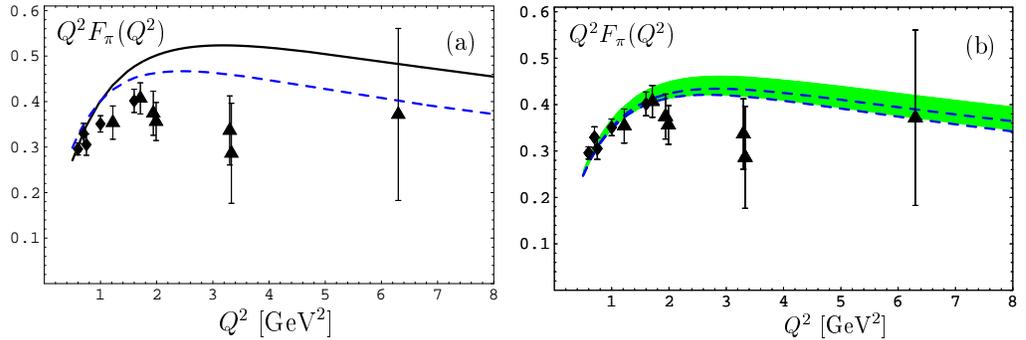}}~~~%
  \caption[*]{(a) Predictions for $Q^2F_\pi(Q^2)$
  obtained with the BMS\ pion DA using standard pQCD within
  the  $\overline{\strut {\rm MS}}$ scheme and adopting
  $\mu_{\rm R}^{2}=Q^2$ (dashed line). The solid line
  corresponds to the $\overline{\strut\text{BLM}}$ scale setting
  introduced in \protect\cite{BPSS04}.
  The experimental data are taken from \protect{\cite{JLAB00}}
  (diamonds) and \protect\cite{FFPI73} (triangles).
  (b) Prediction for $Q^2F_{\pi}(Q^2)$ calculated with the
  ``Maximally Analytic '' procedure and with the BMS ``bunch''  of
  pion DAs.
\label{fig:softFF}}
\end{figure}

The next step is to apply for the calculation of $F_{\pi}^{\rm
Fact}(Q^{2})$ APT. This is done by employing two different
analytization procedures: (i) A {\it Maximally Analytic}
prescription \cite{BPSS04}, meaning that analyticity has been
imposed not only on the coupling, but also on its powers, which,
therefore, have their own dispersive images. This amounts to
\begin{equation}
  \left[F_{\pi}^{\rm Fact}(Q^2; \mu_{\rm R}^{2})\right]_{\rm MaxAn}
 \ =\bar{\alpha}_{\rm s}^{(2)}(\mu_{\rm R}^{2})\,
    {\cal F}_{\pi}^{\rm LO}(Q^2)
   + \frac{1}{\pi}\,
      {\cal A}_{2}^{(2)}(\mu_{\rm R}^{2})\,
       {\cal F}_{\pi}^{\rm NLO}(Q^2;\mu_{\rm R}^{2})\,,
\label{eq:pffMaxAn}
\end{equation}
where $\bar{\alpha}_{\rm s}^{(2)}(\mu_{\rm R}^{2})$ is the 
two-loop analytic coupling and ${\cal A}_{2}^{(2)}(\mu_{\rm
R}^{2})$ the analytic version of its second power in two-loop
order \cite{BPSS04}. (ii) Another procedure, we call \cite{BPSS04}
{\it Naive Analytic}, replaces in $F_{\pi}^{\rm Fact}$ the strong
coupling and its powers by the analytic coupling
$\bar{\alpha}_{\rm s}$ and its powers $\left[\bar{\alpha}_{\rm
s}^{(2)}(\mu_{\rm R}^{2})\right]^2$, entailing the requirement
\cite{SSK99}
\begin{equation}
  \left[F_{\pi}^{\rm Fact}(Q^2; \mu_{\rm R}^{2})\right]_{\rm NaivAn}
\ = \bar{\alpha}_{\rm s}^{(2)}(\mu_{\rm R}^{2})\,
    {\cal F}_{\pi}^{\rm LO}(Q^2)
   + \frac{1}{\pi}\,
      \left[\bar{\alpha}_{\rm s}^{(2)}(\mu_{\rm R}^{2})\right]^2\,
       {\cal F}_{\pi}^{\rm NLO}(Q^2;\mu_{\rm R}^{2})\,.
\label{eq:pffNaivAn}
\end{equation}
The results for $F_\pi$ vs.\ the experimental data are displayed
in Fig.\ 3(b) and Fig.\ 4.
\begin{figure}[!thb]
 \centerline{\includegraphics[width=0.86\textwidth]{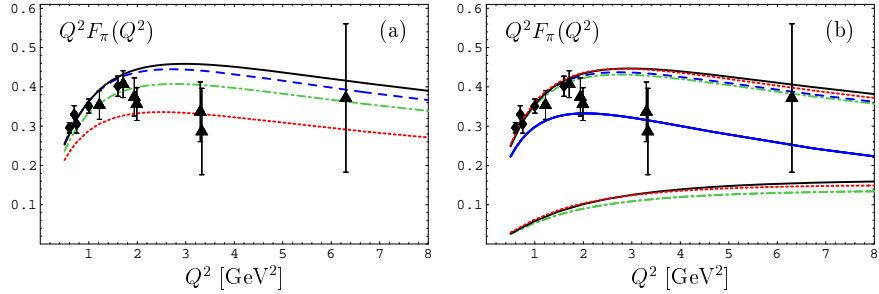}}
  \caption[*]{Predictions for $Q^2F_\pi(Q^2)$ using APT and the
  BMS\ DA in conjunction with the ``Naive Analytic'' (a) and
  ``Maximally Analytic'' (b) procedures:
  $\overline{\strut \text{MS}}$ scheme and $\mu_{\rm R}^{2}=Q^2$
  (dashed line);
  BLM\ (dotted line); $\overline{\strut{\rm BLM}}$ (solid line);
  $\alpha_V$-scheme (dash-dotted line). The single solid line in panel
  (b) shows the prediction for the soft form-factor part;
  below this, the corresponding hard contributions are displayed.
  \label{fig:pidatasum}}
\end{figure}

\section{Summary and conclusions}
The BMS pion DAs \cite{BMS01} successfully pass the comparison
with the CLEO data \cite{CLEO98} at the $1\sigma$ level, as
highlighted in Fig.\ 2 (conforming also with the CELLO data
\cite{CELLO91}). Employing 2-loop APT---naive and maximal--- we
have calculated the hard part of the electromagnetic pion form
factor within various renormalization schemes and using different
scale settings. Joining the hard part with the soft one on the
basis of local duality, we have derived predictions that reproduce
the available data rather well, especially using the ``Maximally
Analytic'' procedure (Fig.\ 3(b)). Moreover, we found that this
procedure minimizes the influence of scheme and scale-setting
ambiguities on the form-factor predictions.

\end{document}